\documentclass[12pt,a4paper,epsf]{article}
\usepackage{graphics}
\usepackage{amssymb,amsmath}
\usepackage[dvips]{lscape,graphicx}
\usepackage{cite}
\usepackage{longtable}

\newcommand{\ct}{\cite}
\newcommand{\lb}{\label}

\newcommand{\bc}{\begin{center}}
\newcommand{\ec}{\end{center}}
\newcommand{\bd}{\begin{displaymath}}
\newcommand{\ed}{\end{displaymath}}
\newcommand{\be}{\begin{equation}}
\newcommand{\ee}{\end{equation}}
\newcommand{\ba}{\begin{array}}
\newcommand{\ea}{\end{array}}
\newcommand{\bea}{\begin{eqnarray}}
\newcommand{\eea}{\end{eqnarray}}
\newcommand{\bt}{\begin{tabular}}
\newcommand{\et}{\end{tabular}}

\newcommand{\bp}{\begin{picture}}
\newcommand{\ep}{\end{picture}}
\newcommand{\bfi}{\begin{figure}}
\newcommand{\efi}{\end{figure}}

\def\fun#1#2{\lower3.6pt\vbox{\baselineskip0pt\lineskip.9pt
\ialign{$\mathsurround=0pt#1\hfil##\hfil$\crcr#2\crcr\sim\crcr}}}

\begin{document}



\vspace{1cm}

\title{\LARGE \bf {Graviweak Unification, Invisible Universe and Dark Energy}}
\author{\large \bf  C.R.~Das${}^{1}$\footnote
{crdas@cftp.ist.utl.pt}\,\,, L.V.~Laperashvili ${}^{2}$
\footnote{laper@itep.ru}\,\,, A.~Tureanu ${}^{3}$
\footnote{anca.tureanu@helsinki.fi}\\[5mm]
\itshape{\large ${}^{1}$ Centre for Theoretical Particle Physics,
CFTP (IST),}\\ {\large \it Avenida Rovisco Pais, 1
1049-001
Lisbon, Portugal}\\
\itshape{\large ${}^{2}$ The Institute of Theoretical and
Experimental Physics,}\\ {\large \it Bolshaya Cheremushkinskaya, 25, 117218 Moscow, Russia}\\
\itshape{\large ${}^{3}$ Department of Physics, University of
Helsinki,}\\ {\large \it P.O. Box 64, FIN-00014 Helsinki, Finland
}}

\date{}

\maketitle

\thispagestyle{empty}

\begin{abstract}

We consider a graviweak unification model with the assumption of
the existence of a hidden (invisible) sector of our Universe,
parallel to the visible world. This Hidden World (HW) is assumed
to be a Mirror World (MW) with broken mirror parity. We start with
a diffeomorphism invariant theory of a gauge field valued in a Lie
algebra  $\mathfrak g$, which is broken spontaneously to the
direct sum of the spacetime Lorentz algebra and the Yang--Mills
algebra: $\tilde {\mathfrak g} = {\mathfrak su}(2)^{(grav)}_L
\oplus {\mathfrak su}(2)_L$ -- in the ordinary world, and $\tilde
{\mathfrak g}' = {{\mathfrak su}(2)'}^{(grav)}_R \oplus {\mathfrak
su}(2)'_R$ -- in the hidden world. Using an extension of the
Plebanski action for general relativity, we recover the actions
for gravity, $SU(2)$ Yang--Mills and Higgs fields in both (visible
and invisible) sectors of the Universe, and also the total action.
After symmetry breaking, all physical constants, including the
Newton's\, constants, cosmological constants, Yang--Mills
couplings, and other parameters, are determined by a single
parameter $g$ present in the initial action, and by the Higgs
VEVs. The dark energy problem of this model predicts  a too large
supersymmetric breaking scale ($M_{SUSY}\sim 10^{10}$ GeV), which
is not within the reach of the LHC experiments.

\end{abstract}



\thispagestyle{empty}

\clearpage\newpage

\section{Introduction}

In this paper, developing the ideas of Ref.~\ct{1a}, we construct
a model of unification of gravity and the $SU(2)$ gauge and Higgs
fields, using only the graviweak part of the model proposed in
Ref. \ct{1}. In contrast to Ref.~\ct{1}, we assume the existence
of the hidden sector of the Universe \ct{2,3,4,5,6}, when this
Hidden World (HW) is a Mirror World (MW) with broken mirror parity
(see Refs.~\ct{7,8,9,10,11,11a} and reviews \ct{12,13,14,15}). We
also discuss the problem of the Dark Energy (DE), and give the
prediction for the supersymmetry breaking scale.

Considering a diffeomorphism invariant theory of a gauge field
with a Lie algebra broken to the direct sum of the spacetime
Lorentz algebra, Yang-Mills algebra and their complement, we
assume that the action, obtained in the graviweak model \cite{1}, has a
modified duplication for the hidden sector of the Universe, in
accordance with ideas proposed in Ref.~\ct{1a}.

The paper is organized as follows. In Section~1 we discuss the
graviweak unification model in the ordinary and hidden worlds of
the Universe. An extension of  Plebanski's action for general
relativity was used in this unification model, and in Section~2 we
introduce the main ideas of the Plebanski's theory of gravity
\ct{17}. In Section~3 we review the theory of the Mirror World
(MW) parallel to the visible Ordinary World (OW), and also the
model of the Mirror World with broken mirror parity (MP), which
assumes that the Higgs VEVs in the O- and M-worlds are not equal
{\footnote {In this paper the superscript 'prime' denotes the M-
or hidden H-world.}: $\langle\phi\rangle=v$,
$\langle\phi'\rangle=v'$ and $v\neq v'$. In Section~4, we build on
the proposal made in Ref.~\ct{1a}, according to which the
OW-gravity is described by the self-dual left-handed Plebanski's
gravitational action, while the gravity in the MW  is described by
the anti-self-dual right-handed gravitational action. Constructing
the graviweak unification model in both sectors of the Universe
\cite{1}, we start with a $\mathfrak g = \mathfrak
{spin}(4,4)$-invariant extended Plebanski's action.  We\ show that
the action for gravity and $SU(2)$ Yang-Mills and Higgs fields,
constructed in the ordinary world, has a modified duplication for
the hidden sector of the Universe. After symmetry breaking of the
graviweak unification, we obtain {Newton'}s constants and the
cosmological constants, which are not equal in the OW and MW. The
mirror cosmological constant is larger than the ordinary
cosmological constant, while the OW and MW Yang-Mills coupling
constants are equal. In Section~5 we discuss the problems of
communications between visible and invisible worlds. These
communications are given by $L_{(mix)}$-term of the total
Lagrangian of the Universe, which describes extremely small
contributions of communication processes.  We also discuss the
experimental tests of the parity violation in gravity. Section~6
is devoted to the problem of the DE. We show that in the framework
of the broken graviweak unification model, the tiny value of the
DE density, given by the recent astrophysical and cosmological
measurements, leads to  the cut-off scale of the theory being less
than the Planck scale, and equal to the large supersymmetry
breaking scale $M_{SUSY}\sim 10^{10}$ GeV, which is not within the
reach of the LHC experiments. However, we discuss a possibility to
reduce the theoretically predicted value of $M_{SUSY}$.

\section{Unification models and  Plebanski's theory of
gravity}

General relativity (GR), originally formulated by Einstein as the
dynamics of a metric, $g_{\mu\nu}$, was presented by Plebanski
\ct{17}, Ashtekar \ct{18,19} and others \ct{20,21,22} in a
self-dual approach, when the true configuration variable of GR
is a connection corresponding to the gauging of the local Lorentz
group, $SO(1,3)$, and the spin group, $Spin(1,3)$ (or its chiral
subgroup).

The idea that interactions are described by a gauge
connection common to both Yang--Mills theory and GR, leads to the
unification of gravity with the other forces. In the unification
model \ct{1}, the fundamental variable is a connection, $A$,
valued in a Lie algebra, $\mathfrak g$, that includes a subalgebra
$\tilde {\mathfrak  g}$:
\be \tilde {\mathfrak  g} = {\mathfrak  g}^{(spacetime)}
\oplus
{\mathfrak g}_{YM}, \lb{1} \ee
which is the direct sum of the Lorentz algebra (or a chiral
subalgebra of it) and a Yang--Mills gauge algebra.

Previously,  graviweak unification models were presented in
Refs.~\ct{23,24,25}. The gravi-GUT unification was suggested in
\ct{26,27,28} and discussed in \ct{29}. As it was shown in Ref.~
\ct{29}, these unifications are based on the formulation of a
fully $\mathfrak  g$-invariant theory via an extension of the
covariant Plebanski action of gravity \ct{17,18,19,20,21,22}. This
kind of the extension previously had been studied in
Refs.~\ct{30,31,32,33,34,35,36,37,38} (see also review \ct{39} and
Ref.~\ct{1a}), especially for the self-dual Plebanski action,
where ${\mathfrak  g}^{(spacetime)} = \mathfrak  {su}(2)_L$.

In Ref.~\ct{1} $, {\mathfrak g}^{(spacetime)}= \mathfrak{
spin}(1,3)$ is the gravitational gauge algebra, and the
Yang--Mills gauge algebra is a spin algebra, ${\mathfrak g}_{YM} =
\mathfrak {spin}(N)$. Finally, the model of unification of
gravity, the $SU(N)$ or $SO(N)$ gauge fields and Higgs bosons is
based on the full initial gauge algebra $\mathfrak g = \mathfrak
{spin}(p,q)$.

The main idea of Plebanski's formulation of the 4-dimensional
theory of gravity \ct{17} is the construction of the gravitational
action from the product of two 2-forms (see \ct{17,18,19,20,21,22}
and \ct{30,31,32,33,34,35,36,37,38,39}). These 2-forms are
constructed using the connection $A^{IJ}$ and tetrads, or frames,
$e^I$ as independent dynamical variables. Both $A^{IJ}$ and $e^I$,
also $A$, are 1-forms:
\be   A^{IJ} =  A_{\mu}^{IJ}dx^{\mu} \quad {\mbox{and}}\quad
       e^I = e_{\mu}^Idx^{\mu},
                          \lb{2} \ee
\be  A = \frac 12 A^{IJ}\gamma_{IJ}. \lb{3} \ee
Here the bivector generators $\gamma_{IJ}$  can be understood as
the product of $Cl(1,3)$ Clifford algebra basis vectors:
$\gamma_{IJ}=\gamma_{I}\gamma_{J}$.

The indices $I,J = 0,1,2,3$ refer to the spacetime with Minkowski
metric $\eta_{IJ}$: $\eta^{IJ} = {\rm diag}(1,-1,-1,-1)$. This is
a flat space which is tangential to the curved space with the
metric $g_{\mu\nu}$. The world interval is represented as $ds^2 =
\eta_{IJ}e^I \otimes e^J$, i.e.
\be g_{\mu\nu} = \eta_{IJ} e^I_{\mu}\otimes e^J_{\nu}.
 \lb{4} \ee
Considering the case of the Minkowski flat spacetime with the
group of symmetry $SO(1,3)$, we have the capital latin indices
$I,J,...=0,1,2,3$, which are vector indices under the rotation
group $SO(1,3)$.

In the general case of the gauge symmetry $\mathfrak G$ with the Lie
algebra $\mathfrak g = spin(p,q)$, we have $I,J = 0,1,2,
...,p+q-1$.

The 2-forms $B^{IJ}$ and $F^{IJ}$ are defined as:
\be
      B^{IJ} = e^I\wedge e^J = \frac 12
      e_{\mu}^Ie_{\nu}^Jdx^{\mu}\wedge dx^{\nu},
                    \lb{5} \ee
\be
      F^{IJ} = \frac 12 F_{\mu\nu}^{IJ}dx^{\mu}\wedge dx^{\nu}.
                    \lb{6} \ee
Here the tensor $F_{\mu\nu}^{IJ}$ is the field strength of the
spin connection $A_{\mu}^{IJ}$:
\be
     F_{\mu\nu}^{IJ} = \partial_{\mu}A_{\nu}^{IJ} -
         \partial_{\nu}A_{\mu}^{IJ} - [A_{\mu}, A_{\nu}]^{IJ},
                               \lb{7} \ee
which determines the Riemann--Cartan curvature:
\be
       R_{\kappa \lambda \mu \nu} = e_{\kappa}^I e_{\lambda}^JF_{\mu\nu}^{IJ}.
                  \lb{8} \ee
We also consider the 2-forms $B$ and $F$:
\be B= \frac 12 B^{IJ}\gamma_{IJ} \quad  {\rm and} \quad  F= \frac
12 F^{IJ}\gamma_{IJ},  \lb{9} \ee
which are $spin(1,3)$ valued 2-form fields, and:
\be F = dA + \frac 12 [A, A].  \lb{10} \ee

In the Plebanski BF-theory, the gravitational action with nonzero
cosmological constant $\Lambda$ is given by the integral:
\be  I_{GR} = \frac{1}{\kappa^2}\int
\epsilon^{IJKL}\left(B^{IJ}\wedge
    F^{KL} + \frac{\Lambda}{4}B^{IJ}\wedge B^{KL}\right),
                                  \lb{11} \ee
where $\kappa^2=8\pi G_N$, $G_N$ is the gravitational constant, $
M_{Pl}^{red.} = 1/{\sqrt{8\pi G_N}}$.

For any antisymmetric tensors $F_{\mu\nu}$ there exist dual
tensors given by the Hodge star dual operation:
\be
 F^*_{\mu\nu}\equiv \frac {1}{2\sqrt{-g}}\epsilon ^{\rho\sigma}_{\mu\nu}F_{\rho\sigma},
                                 \lb{12} \ee
and for any antisymmetric tensors $A^{IJ}$ there exists dual
operation:
\be  A^{\star IJ} = \frac 12 \epsilon^{IJKL}A^{KL}.
                                 \lb{13} \ee
Here $\epsilon$ is the completely antisymmetric tensor with
$\epsilon^{0123} = 1$.

We can define the algebraic self-dual and anti-self-dual
components of $A^{IJ}$:
\be A^{(\pm)\,IJ}=({\cal P}^{\pm}A)^{IJ} = \frac 12 (A^{IJ} \pm
iA^{\star\,IJ}).
                                 \lb{14} \ee
The two projectors ${\cal P}^{\pm}= \frac 12(\delta^{IJ}_{KL} \pm
\epsilon^{IJ}_{KL})$ realize explicitly the familiar homomorphism:
\be
   \mathfrak{so}(1,3) = \mathfrak{su}(2)_R \oplus
   \mathfrak{su}(2)_L,
                                 \lb{15} \ee
which rather than self-dual and anti-self-dual are more commonly
dubbed right-handed and left-handed.

To make the mapping more explicit, it is convenient to pick out
the time direction $I = 0$, and define
\be
                 A^{(\pm)i} = A^{(\pm) 0i},
                     \lb{16} \ee
with $\large i = 1,2,3$ as an $SU(2)_L^{(grav)}$ adjoint index.

(Anti)self-duality then means:
 \be  A^{(\pm)0i} = \pm \frac i2 \epsilon^i_{jk}A^{(\pm) jk}.
                     \lb{17} \ee
The correct gauge was chosen by Plebanski, when he introduced in
the gravitational action the Lagrange multipliers $\psi_{ij}$ --
an auxiliary fields, symmetric and traceless. These auxiliary
fields $\psi_{ij}$ provide the correct number of constraints.

Including the constraints, we obtain the following gravitational
action:
\be I(\Sigma,A,\psi) = \frac{1}{\kappa^2} \int [\Sigma^i\wedge F^i
+
 (\Psi^{-1})_{ij}\Sigma^i\wedge \Sigma^j]
                      \lb{18} \ee
with
\be   (\Psi^{-1})_{ij} = \psi_{ij} - \frac{\Lambda}{6}\delta_{ij}.
                     \lb{19} \ee
Using the simpler self-dual variables instead of the full Lorentz
group, Plebanski \ct{17} and the authors of
Refs.~\ct{18,19,20,21,22} suggested to consider the left-handed
$\mathfrak{su}(2)_L$-invariant gravitational action (\ref{18})
with self-dual $F=F^{(+)i}$ and $\Sigma=\Sigma^{(+)i}$. In
general, we have:
\be  \Sigma^{(\pm) i} = e^0\wedge e^i \pm i\frac 12
     \epsilon^i_{jk}e^j\wedge e^k.
                          \lb{20} \ee
The self-dual action (\ref{18}) is equivalent to the
Einstein-Hilbert action for general relativity with cosmological
constant $\Lambda$.

\section{Invisible Universe}

\subsection{Mirror World}

Mirror matter is a new form of matter which is predicted to exist
if mirror symmetry is respected by Nature. At the present time,
evidence that mirror matter actually exists in the Universe is in
abundance, coming from a range of observations and experiments in
astronomy, particle physics, stars and planetary science.

The results of Refs.~\ct{2,3,4,5,6,7,8,9,10,11,11a,12,13,14,15}
are based on the hypothesis of the existence in Nature of the
invisible mirror parallel to the visible ordinary world. The
Standard Model (SM) group of symmetry $G_{SM}$ was enlarged to
$G_{SM}\times G'_{SM'}$, where $G_{SM}$ stands for the observable
SM, while $G'_{SM'}$ is its mirror gauge counterpart. The
M-particles are singlets of $G_{SM}$ and the O-particles are
singlets of $G'_{SM'}$. {\it These different O- and M-worlds are
coupled only by gravity, or possibly by another very weak
interaction} \ct{2,3,4,5,6}.

The M-world is a mirror copy of the O-world and contains the same
particles and types of interactions as our visible world. The
observable elementary particles of our O-world have the
left-handed (V-A) weak interactions which violate P-parity. If a
hidden  M-world exists, then mirror particles participate in
the right-handed (V+A) weak interactions and have the opposite
chirality. Lee and Yang were the first \ct{2} to suggest such a
duplication of the worlds, which restores the left-right symmetry
of Nature. They introduced a concept of right-handed particles,
but their R-world was not hidden. The term "Mirror Matter" was
introduced by Kobzarev, Okun and Pomeranchuk \ct{3}. They first
suggested the MW as the hidden (invisible) sector of
our Universe, which interacts with the ordinary (visible) world
only via gravity, or another very weak interaction. They have
investigated a variety of phenomenological implications of such
parallel worlds (see reviews \ct{12,13}).

Superstring theory also predicts that there may exist in the
Universe another form of matter -- hidden ('shadow') matter, which
only interacts with ordinary matter via gravity or
gravitational-strength interactions \ct{4,5,6}.  According to the
superstring theory, the two worlds, ordinary and shadow, can be
viewed as parallel branes in a higher dimensional space, where
O-particles are localized on one brane and hidden particles -- on
another brane, and gravity propagates in the bulk. In
Refs.~\ct{40,41,42,43,44,45} we considered the theory of the
superstring-inspired $E_6$ unification with different types of
breaking in the visible (O) and hidden (M) worlds.

\subsection{Mirror world with broken mirror parity}

If the ordinary and mirror worlds are identical, then O- and
M-particles should have the same cosmological densities. But this
is immediately in conflict with recent astrophysical measurements.

Astrophysical and cosmological observations (see, for example,
\ct{46,47,48}) have revealed the existence of Dark Matter (DM)
which constitutes about 23\% of the total energy density of the
present Universe. This is five times larger than all the visible
matter, $\Omega_{DM}: \Omega_{M} \simeq 5 : 1$. In parallel to the
visible world, the mirror world conserves mirror baryon number and
thus protects the stability of the lightest mirror nucleon. Mirror
particles have been suggested as candidates for the inferred dark
matter in the Universe (see
Refs.~\ct{7,8,9,10,11,11a,12,13,14,15}). This theory explains the
right amount of dark matter, which is generated via the mirror
leptogenesis \ct{49,50a,50b,50c,51}, just like the visible matter
is generated via ordinary leptogenesis \ct{52,53}.

Therefore, mirror parity (MP) is not conserved, and the O- and M-
worlds are not identical. In
Refs.~\ct{7,8,9,10,11} it was suggested that the VEVs of the Higgs
doublets $\phi$ and $\phi'$ are not equal:
\be \langle\phi\rangle=v,\quad \langle\phi'\rangle=v' \quad {\rm {and}}\quad v\neq v'.
                          \lb{39} \ee
Introducing the parameter characterizing the violation of MP,
\be  \zeta = \frac {v'}{v} \gg 1,
                          \lb{40} \ee
we have the estimates of Refs.~\ct{7,8,9,10,11} and
\ct{54,55,55a,55b}:
\be  \zeta > 30,\quad \zeta \sim 100.
                          \lb{41} \ee
Then the masses of mirror fermions and massive bosons are scaled
up by the factor $\zeta$ with respect to the masses of their
counterparts in the OW:
\be m'_{q',l'} = \zeta m_{q,l},
                          \lb{42} \ee
and \be
   M'_{W',Z',\phi'} = \zeta  M_{W,Z,\phi}, \lb{43} \ee
while photons and gluons remain massless in both worlds.

In the language of neutrino physics, the O-neutrinos
$\nu_e,\,\,\nu_{\mu},\,\,\nu_{\tau}$ are active neutrinos, and the
M-neutrinos $\nu'_e,\,\,\nu'_{\mu},\,\,\nu'_{\tau}$ are sterile
neutrinos \ct{56}. If MP is conserved ($\zeta = 1$), then the
neutrinos of the two sectors are strongly mixed (see Refs.~
\ct{7,8,9,10,11}). However, the present experimental and
cosmological limits on the active-sterile neutrino mixing do not
confirm this result.

In Refs.~{\ct{56a,56b}} the exact parity symmetry explains the
solar neutrino deficit, the atmospheric neutrino anomaly and the
LSND experiment.

In the context of the SM, in addition to the fermions with
non-zero gauge charges, one introduces also the gauge singlets,
the so-called right-handed neutrinos $N_a$ with large Majorana
mass terms. They have equal masses in the O- and M-worlds
\ct{7,8}:
\be M'_{\nu,a} = M_{\nu,a}. \lb{44} \ee
According to the usual seesaw mechanism \ct{52,53}, heavy
right-handed neutrinos are created at the seesaw scale $M_R$ in
the O-world and $M'_R$ in the M-world. From the Lagrangians,
considering the identical O- and M- Yukawa couplings, we obtain:
\be m_{\nu}^{(')}=\frac {{v^{(')}}^2}{M_R^{(')}}. \lb{45} \ee
The equality of the seesaw scales, $M'_R = M_R$, immediately leads
to the relation between the masses of light neutrinos:
\be
  m'_{\nu} = \zeta^2 m_{\nu}. \lb{46} \ee
Here sterile neutrinos are $\sim 10^4$ times heavier than their
O-partners. Eqs.~(\ref{45}) and (\ref{46}) were first obtained in
Ref.~\ct{54}.

\section{Graviweak action in the visible and invisible sectors of
the Universe}

In Ref.~\ct{1a} we suggested to describe the gravity in the
visible Universe by the self-dual left-handed Plebanski's
gravitational action, while the gravity in the invisible Universe
-- by the anti-self-dual right-handed gravitational action:
\be I^{(')}_{(gravity)}(\Sigma^{(')},A^{(')},\psi^{(')} ) =
   \frac {1}{{\kappa^{(')}}^2} \int [{\Sigma^{(')}}^i\wedge {F^{(')}}^i +
 ({\Psi^{(')}}^{-1})_{ij}{\Sigma^{(')}}^i\wedge {\Sigma^{(')}}^j],
                      \lb{21} \ee
where $A,\Sigma = A^{(+)},\Sigma^{(+)}$ are self-dual
(left-handed) fields in the OW, and $A',\Sigma' =
A^{(-)},\Sigma^{(-)}$ are anti-self-dual (right-handed) fields in
the MW, and $\Sigma^{(\pm)}$ are given by Eq.~(\ref{20}).

Developing the ideas of Ref.~\ct{1a}, we consider the graviweak
unification model in both sectors of the Universe, visible and
invisible. We start with a $\mathfrak g = \mathfrak
{spin}(4,4)$-invariant extended Plebanski's action, as proposed in
Ref.~\ct{1}:
\be I(A, B, \Phi) = \frac{1}{g} \int_{\mathfrak M}\langle BF +
B\Phi B + \frac 13 B\Phi \Phi \Phi B \rangle. \lb{22} \ee
The wedge product $\langle...\rangle$ is assumed between the
forms. In this action with a parameter $g$, the connection, $A =
A^{IJ}\gamma_{IJ}$, is the independent physical variable
describing the geometry of the spacetime, while $\Phi$, or
$\Phi_{IJKL}$, are auxiliary fields. Here, $ F = dA + \frac 12 [A,
A]$ is the curvature and $B=B^{IJ}\gamma_{IJ}$ is a
$\mathfrak{spin}(4,4)$-valued 2-form auxiliary field. The
generators $\gamma_{IJ}=\gamma_{I}\gamma_{J}$ of the
$\mathfrak{spin}(4,4)$-group have indices running over all
$8\times 8$ values: $I,J = 0,1,2,...,6,7$.

Now we distinguish the following two actions:

1) the $\mathfrak {spin}(4,4)_L$-invariant action $I_{left}(A, B,
\Phi)$ with self-dual left-handed fields \\ $A=A^{(+)}$,
$B=B^{(+)}$ and $\Phi_{IJKL}$ -- in the ordinary (visible) world
OW, and

2) the $\mathfrak {spin}(4,4)_R$-invariant action $I_{right}(A',
B', \Phi')$ with anti-self-dual right-handed fields $A'=A^{(-)}$,
$B'=B^{(-)}$ and $\Phi'_{IJKL}$ -- in the hidden (invisible) world
MW.

The action (\ref{22}) is a modification of  Plebanski's
action, which allows the symmetry breaking to a nontrivial vacuum
expectation value (VEV).

Varying the fields $A,B$ and $\Phi$, one obtains the field equations:
\be  {\cal D}B = dB + [A,B] = 0,  \lb{23} \ee
where ${\cal D}$ is the covariant derivative, ${\cal D}_{\mu}^{IJ}
= \delta^{IJ}\partial_{\mu} - A_{\mu}^{IJ}$,
and%
\be   F = -2\left(\Phi +\frac 13\Phi \Phi \Phi\right)B,  \lb{24} \ee
\be    B^{IJ} B^{KL} = - \frac 1{16} B^{IJ}
\Phi^{KL}_{MN}\Phi^{MN}_{PQ}B^{PQ}. \lb{25} \ee
The first equation (\ref{23}) describes the dynamics, while
(\ref{24}) and (\ref{25}) determine $B$ and $\Phi$.

Solving the equations of motion for $B$ in terms of $F$, we obtain
the following $\mathfrak g$-invariant gravitational, gauge and
Higgs field action:
\be
   I(e,A) = \frac{3}{8g}\int_{\mathfrak M} \langle F F^{\star} \rangle.
   \lb{26} \ee
where  $ F = dA + \frac 12 [A, A]$ and $A = A^{IJ}\gamma_{IJ}$
with $I,J = 0,1,2,...,6,7$.

Eqs.~(\ref{23})-(\ref{26}) are valid in the OW, and  similar
equations hold in MW for $A',B',F'$ and $\Phi'$.

For completeness, we briefly present, according to Ref. \cite{1},
the spontaneous symmetry breaking of the $\mathfrak
g$-invariant action (\ref{26}) that produces the dynamics of the $
SU(2)_L$-gravity, and the $SU(2)_L$ gauge and Higgs fields with
subalgebra
$$  \tilde {\mathfrak g} = {{\mathfrak su}(2)}^{(grav)}_L
\oplus {\mathfrak su}(2)_L.$$ The indices $\large
a, b \in \{0,1,2,3\}$ are used to sum over a subset of
$I, J \in {0,1,2, ...,7}$, and thereby select a $\mathfrak
{spin}(1,3)$ subalgebra of $\mathfrak {spin}(4,4)$. The indices
$ m, n \in \{4,5,6,7\}$ sum over the rest. We also
consider  $\large i, j \in \{1,2,3\}$, thus selecting a $\mathfrak
{su}(2)_L^{grav}$ subalgebra of $\mathfrak {spin}(4,4)_L$.

Analogous equations are valid in the MW with the initial
$\mathfrak {spin}(4,4)_R$-algebra, and with a subalgebra:
\be  \tilde {\mathfrak g}' = {{\mathfrak su}(2)'}^{(grav)}_R
\oplus {\mathfrak su}(2)'_R.      \lb{27} \ee
The  spontaneous symmetry breaking of the graviweak unification in this model
gives separate parts of the connection in terms of the following
2-forms:
\be   A = \frac 12 \omega + \frac 14 E + A_W. \lb{28} \ee
Here the gravitational spin connection is:
\be   \omega = \omega^{ab}\gamma_{ab}, \lb{33} \ee
or
\be  \omega =  \omega^i \sigma_i, \lb{34} \ee
which is valued in $\mathfrak {su}(2)_L^{(grav)}$, and $\sigma_i$
are Pauli matrices, $i=1,2,3$.

The frame-Higgs connection
\be   E = E^{am}\gamma_{am},   \lb{35} \ee
which is valued in the off-diagonal complement of $\mathfrak
{spin}(4,4)$, is assumed to have the expression:
\be  \ E = e\phi = e^a_{\mu}\sigma_a\phi^i\sigma_i dx^{\mu}.
                                          \lb{36} \ee
The field $\phi=\phi^i\sigma_i$ is the scalar Higgs doublet of
$\mathfrak su(2)_L$.

The gauge field:
\be  A_W = \frac 12 A^{mn}\gamma_{mn}, \lb{37} \ee
or
\be   A_W = \frac 12 A_W^i \tau_i, \lb{38} \ee
is valued in $\mathfrak su(2)_L$.

Analogous equations exist in MW for $A',\omega', E', A'_{W'}$.

Finally, using the results of Refs.~\ct{1a} and \ct{1}, we have the
following actions for gravitational, the $SU(2)_L$ and $SU(2)'_R$
gauge and Higgs fields in the ordinary and hidden sectors of the
Universe, respectively:
$$  I^{(')}(e^{(')},\phi^{(')},A^{(')},A^{(')}_{W^{(')}})= \frac{3}{8g}
\int_{\mathfrak M} d^4x|e^{(')}|\Big(-\frac 1{16}{|\phi^{(')}|}^2
R^{(')} + \frac{3}{32}{|\phi^{(')}|}^4 + \frac
1{16}{{R^{(')}}_{ab}}^{cd} {{R^{(')}}^{ab}}_{cd}$$ \be - \frac 12
{\cal D}_a{\phi^{(')}}^{\dagger} {\cal D}^a\phi^{(')} - \frac 14
{F^{(')}}^{mn}_{W^{(')},ab}{F^{(')}}^{mn,ab}_{W^{(')}} \Big).
\lb{51} \ee
Here $g=g'$, $R^{(')}$ is the Riemann curvature scalar,  $|\phi|^2
=  {\phi}^{\dag}\phi$ is the squared magnitude of
the Higgs field, ${\cal D} \phi = d\phi + [A_W,\phi]$ is the
covariant derivative of the Higgs, and $F_W = dA_W + [A_W,A_W]$ is
the curvature of the gauge field. Similar notations are made for
$A',\phi',A'_{W'}$. The third term of (\ref{51}) is a modification
to the standard gravitation related to the Gauss--Bonnet
topological action \ct{57,58}.

The nontrivial vacuum solutions to the actions (\ref{51}) give the
non-vanishing Higgs vacuum expectation values (VEVs):
$v^{(')}=\langle\phi^{(')}\rangle=\phi^{(')}_0$, at which the
standard Higgs potentials have an extremum corresponding to a de
Sitter spacetime background solution:
\be {(v^{(')})}^2 = \frac{R^{(')}_0}{3}   \lb{52a} \ee
(see details in Ref.~ \ct{1}). Here $R^{(')}_0\neq 0$ is a
constant background scalar curvature. We see that in this model
the spacetime geometry determines the Higgs VEVs, e.g. the masses
of the SM and SM' particles.

After the symmetry breaking of the graviweak unification in both
worlds, we obtain:

1) the {Newton'}s constants in the OW and MW of our Universe,
respectively, are equal to
\be G^{(')}_N = \frac{128g}{3{v^{(')}}^2}, \lb{52} \ee

2) the  cosmological constants $\Lambda^{(')}$  (given by the
second term of (\ref{51})) are equal to
\be \Lambda^{(')} = \frac 34 {v^{(')}}^2,  \lb{53} \ee

3) the weak coupling constants:

\be   {g^{(')}}^2_W=\frac{8g}{3}.  \lb{54} \ee
Finally, we have the following relations:
\be G'_N = \frac{G_N}{\zeta^2}, \quad
  \Lambda' = \zeta^2 \Lambda, \quad M_{Pl}' = \zeta M_{Pl},
     \quad       g'_W = g_W.  \lb{55} \ee
All physical constants are determined by a parameter $g$ and the
Higgs VEVs $v^{(')}$. Here we assume that the equality $g'=g$ is a
consequence of the existence of the Grand Unification at the early
stage of the Universe when the mirror parity was unbroken.

Of course, the present model is too simple to have the correct
phenomenological applications. It is necessary to note that all
parameters -- Newton's constants, the cosmological constants, the
gauge couplings $g_{YM}=g_W$, etc., considered in (\ref{51}) --
are the bare parameters, which refer to the Planck scale (see, for
example, the Planck scale physics in Refs.~\ct{59,60,60a}).

Using the experimentally known values of $G_N$, where
$M_{Pl}^{red.} = 1/{\sqrt{8\pi G_N}}\approx 2.43\cdot 10^{18}$
GeV, and $v\approx 246$ GeV, we can obtain the value of $g$ from
Eq.~(\ref{52}). However, we cannot relate the value $g^2_W=8g/3$
with the value of $g_2^2$ obtained by the extrapolation of
experimental values of running $\alpha_2=g_2^2/4\pi $ from the
electroweak scale to the Planck scale. The reason is that such a
running $\alpha_2$ includes the electroweak breaking symmetry and
also is determined by the fermion (quark) contributions, which are
absent in the present model. Nevertheless, the explicit relation
between the gravitational and Yang--Mills couplings is a
consequence of a genuine unification of gravity and Yang--Mills
theory.

\section{Communications between Visible and Hidden worlds}

The broken MP means that parity violation applies not only for the weak
interaction, but also in the gravitational sector. The parity
violation of gravity was considered in Ref.~\ct{61}, where the
authors suggested to find the effect of birefringence amplitude of
gravitational waves, whereby left and right circularly-polarized
waves propagate at the same speed but with different amplitude
evolution. A test of this effect was proposed through
coincident observations of gravitational waves and short gamma-ray
bursts from binary mergers involving neutron stars. Such
gravitational waves are highly left or right circularly-polarized
due to the geometry of the merger. All sky gamma-ray telescopes
can be sensitive to the propagating sector of gravitational parity
violation.

The dynamics of the two worlds of our Universe, visible and
hidden, is governed by the following action (see \ct{1a}):
\be   I = \int [ L_{(grav)} + L'_{(grav)} + L_{SM} + L'_{SM'} +
L_{(mix)}]|e|d^4x,          \lb{56} \ee
where $L_{(grav)}$ is the gravitational (left-handed) Lagrangian
in the visible world, and $L'_{(grav)}$ is the gravitational
right-handed Lagrangian in the hidden world, $L_{SM}$ and
$L'_{SM'}$ are the Standard Model Lagrangians in the O- and
M-worlds, respectively, $L_{(mix)}$ is the Lagrangian describing
all mixing terms (see \ct{7,8,9,10,11,11a,12}) giving small
contributions to physical processes: mirror particles have not
been seen so far, and the communication between visible and hidden
worlds is hard.

There are several fundamental ways by which the hidden world can
communicate with our visible world. The Lagrangian $L_{(mix)}$
describes all possible mixing terms which are consistent with the
symmetries of the theory:
\be L_{mix} = \alpha F^i\wedge {F'}^i  + \frac{\epsilon_Y}{2}
F_{Y,\mu\nu}{F'_{Y}}^{\mu\nu} + \lambda_2\phi^{\dagger} \phi
{\phi'}^{\dagger} \phi', \lb{57} \ee
where the first term is the gravitational mixing term (see
\ct{39}), $F_{Y,\mu\nu}$ (${F'_{Y}}^{\mu\nu}$) is the $U(1)_Y$
($U(1)'_Y$) field strength tensor and $\phi(\phi')$ is the
ordinary (mirror) Higgs doublet (see \ct{62}). The last term of
Eq.~(\ref{57}) provides the 'Higgs-mirror Higgs' mass mixing that
can lead to non-standard Higgs boson physics at the LHC
\ct{62}.

The left-handed gravity interacts not only with visible matter,
but also with mirror matter. The right-handed gravity also
interacts with matter and mirror matter. We assume that a fraction
of the mirror matter exists in the form of mirror galaxies, mirror
stars, mirror planets etc., (see, for example, Ref.~\ct{63}).
These objects can be detected using gravitational microlensing
\ct{64}.

There exists the kinetic mixing between the electromagnetic field
strength tensors for visible and mirror photons (see
Refs.~\ct{65,66} and \ct{66a}):
\be         L_{\gamma}^{{mix}} = \frac{\epsilon_{\gamma}}{2}
F_{\gamma\,\mu\nu}{F'_{\gamma}}^{\mu\nu}.  \lb{58} \ee
The 'photon--mirror photon' mixing induces the
'orthopositronium--mirror orthopositronium' oscillations
\ct{65,66}. Future experiments with orthopositronium were
suggested in Ref.~\ct{67}.

Interactions between visible and mirror quarks and leptons are
expected to take place. Mirror neutrons can oscillate to ordinary
neutrons giving  'neutron--mirror neutron' oscillations
(Refs.~\ct{68,69,70,71}).

Mass mixings between visible and mirror (sterile) neutrinos lead
to the 'neutrino--mirror neutrino' oscillations what was suggested
by Ref.~\ct{54} (see also \ct{72,73} and \ct{12}).

Heavy Majorana neutrinos $N_a$ are singlets of $G_{SM}$ and
$G'_{SM'}$, and they can be messengers between visible and hidden
worlds (see Refs.~\ct{49,50a,50b,50c,51}).

Also any weakly interacting singlet scalar field can be a
messenger between OW and MW.

The search for mirror particles at the LHC is discussed in
Ref.~\ct{74}.

\section{Dark Energy of the Universe}

In both worlds, O and M, the Einstein equations
\be R^{(')}_{\mu\nu} - \frac{1}{2}R^{(')}g^{(')}_{\mu\nu}= 8\pi
G^{(')}_NT^{(')}_{\mu\nu} - \Lambda^{(')} g^{(')}_{\mu\nu}
                      \lb{63} \ee
contain the energy momentum tensor of matter $T^{(')}_{\mu\nu}$.

In our theory the dark energy density of the Universe is:
\be  \rho_{DE} = \rho_{vac} = \frac{\Lambda_{eff}}{8\pi G_N}  +
\frac{\Lambda'_{eff}}{8\pi G'_N},
                        \lb{64} \ee
where
\be   \frac{\Lambda^{(')}_{eff}}{8\pi G^{(')}_N} =
\frac{\Lambda^{(')}}{8\pi G^{(')}_N} + \rho_{vac}^{(SM^{(')})},
                        \lb{64a} \ee
and $\Lambda^{(')}$ are bare cosmological constants (\ref{53}) of
our theory.

Astrophysical measurements \ct{46,47,48} give:
\be  \rho_{DE} \approx 0.73 \rho_{tot} \approx (2.3\times
10^{-3}\,\,{\mbox{eV}})^4.
                        \lb{65} \ee
Recalling that $M^{red}_{Pl} = (8\pi G_N)^{-1}$ GeV and $\large
v'=\zeta v$, we obtain:
\be \rho_{DE} = \rho_{vac} = \frac 34 v^2 (1 +
    \zeta^4)({M^{red.}_{Pl}})^2 + \rho_{vac}^{(SM)} +
    \rho_{vac}^{(SM')}.       \lb{66} \ee
All quantum fluctuations of the matter contribute to the vacuum
energy density $\large \rho_{vac}$ of the Universe. We expect that
quantum field theory is valid up to some cut-off scale $\large
M_{(cut-off)}$, and the vacuum energy density in the  SM and SM$'$
is evaluated by the sum of zero-point energies of quantum fields.
Then we have:
\be \rho_{DE} = \rho_{vac} = \frac 34 v^2 (1 +
    \zeta^4)({M^{red.}_{Pl}})^2 + C(M_{(cut-off)})^4  +
    C'(M'_{(cut-off)})^4.
                                                \lb{67} \ee
The tiny value of the dark energy density $\rho_{DE} \approx
(2.3\times 10^{-3}\,\,{\mbox{eV}})^4$, verified by astronomical
and cosmological observations, leads to the conclusion that
\be \frac 34 v^2 (1 +
    \zeta^4)({M^{red.}_{Pl}})^2 \sim - CM^4_{(cut-off)} -
    C'{M'}^4_{(cut-off)}.
                                 \lb{68} \ee
This means that all matter quantum fluctuations in SM and SM' must
be almost compensated by the contribution of cosmological
constants $\Lambda$ and $\Lambda'$. This compensation is possible
if $C,C'<0$, what means the dominance of degrees of freedom (DOF)
of fermions in the the sum of zero-point energies of quantum
fields.

Assuming $C_1^{(')} = {|-C^{(')}|}^{1/4} \sim 1$, we have:
\be \frac 34 v^2 (1 +
    \zeta^4)({M^{red.}_{Pl}})^2 \sim M^4_{(cut-off)} + {M'}^4_{(cut-off)}.
                                 \lb{69} \ee
According to Eq.~(\ref{55}), we have:
\be M'_{(cut-off)}=\zeta M_{(cut-off)}, \lb{70} \ee
what means that
\be \frac 34 v^2 (1 +
    \zeta^4)({M^{red.}_{Pl}})^2 \sim (1 + \zeta^4)M^4_{(cut-off)}.
                                 \lb{71} \ee
Using $v\approx 246$ GeV and $M^{red.}_{Pl}\approx 2.43\cdot
10^{18}$ GeV, we obtain:
\be M_{(cut-off)}\sim \sqrt{vM^{red.}_{Pl}}\sim  2.3\cdot
10^{10}\,\, {\rm{GeV}}.
                                 \lb{72} \ee
This result means that in the framework of graviweak unification,
the cut-off scale is less than Planck scale.

The existence of supersymmetry can explain the tiny value of
$\rho_{DE}$. If the supersymmetry breaking scale $M_{SUSY}$
coincides with the cut-off scale (\ref{72}), we obtain:
\be M_{SUSY}\sim  10^{10}\,\, {\rm{GeV}}.
                                 \lb{73} \ee
We see that the supersymmetry breaking scale is essentially large,
and not within the reach of the LHC experiments. The next
possibility to reduce the value $M_{SUSY}$, given by
Eq.~(\ref{73}), is to assume the existence of large negative
contributions to $\Lambda^{(')}$, or to assume that $C_1^{(')}\gg
1$. Unfortunately, both these assumptions are doubtful, and we
hope that the future LHC result for $M_{SUSY}$ will shed light on
these issues.

\section{Summary and Conclusions}

In this paper we developed a graviweak unification of gravity and
the SU(2) gauge and Higgs fields, assuming the existence of the
hidden sector of the Universe.

Developing ideas of Refs.~\ct{1a,1}, we presented a graviweak
unification model in the visible and invisible parts of the
Universe. We started with an extended $\mathfrak
g=\mathfrak{spin}(4,4)_L$-invariant Plebanski action in the OW,
and $\mathfrak g=\mathfrak{spin}(4,4)_R$-invariant Plebanski
action in the MW.

We showed that the graviweak symmetry breaking leads to the
following subalgebras: $\tilde {\mathfrak g} = {\mathfrak
su}(2)^{(grav)}_L \oplus {\mathfrak su}(2)_L$ -- in the ordinary
world, and $\tilde {\mathfrak g}' = {{\mathfrak
su}(2)'}^{(grav)}_R \oplus {\mathfrak su}(2)'_R$ -- in the hidden
world. These subalgebras contain the self-dual left-handed gravity
in the OW, and the anti-self-dual right-handed gravity in the MW.

The nontrivial vacuum solutions corresponding to the obtained
actions are non-vanishing Higgs vacuum expectation values (VEVs):
$v^{(')}=\langle\phi^{(')}\rangle=\phi^{(')}_0$.

Using recent  astrophysical and cosmological measurements, we
considered a model of the Mirror World with broken mirror parity
(MP), which assumes that the Higgs VEVs in the visible and
invisible worlds are not equal: $\langle\phi\rangle=v, \quad
\langle\phi'\rangle=v' \quad
{\rm {and}}\quad v\neq v'$. Introducing the parameter
characterizing the violation of MP, $\zeta = \frac {v'}{v} \gg 1$,
we have used the estimate $\zeta \sim 100$.

In this model, we showed that the action for gravity, and the
$SU(2)$ Yang--Mills and Higgs fields, constructed in the ordinary
world, has a modified duplication for the hidden sector of the
Universe. Considering the graviweak unification in the both worlds
of the Universe, we obtained, after symmetry breaking, {Newton'}s
constants $G^{(')}_N = \frac{128g}{3{v^{(')}}^2}$,  and the
cosmological constants $\Lambda^{(')} = \frac 34 {v^{(')}}^2$,
which are not equal in the OW and MW. The mirror cosmological
constant $\Lambda'= \zeta^{2} \Lambda$ is larger than the ordinary
cosmological constant, while the OW and MW Yang--Mills coupling
constants are equal: ${g^{(')}}^2_{YM} = g_W^2 = \frac{8g}{3}$.

We also discussed the problems of communications between visible
and invisible worlds. Mirror particles have not been seen so far
in the visible world, and the communication between visible and
hidden worlds is hard. This communication is given by the
$L_{(mix)}$-term of the total Lagrangian of the Universe, and the
contributions of processes described by $L_{(mix)}$ are extremely
small.

Finally, we considered the problem of the DE. The recent
astrophysical and cosmological measurements give a tiny value of
the dark energy density $\rho_{DE} = \rho_{vac}\simeq (2.3\times
10^{-3}\,\,{\mbox{eV}})^4$, what means that the sum
$\frac{\Lambda}{8\pi G_N}  + \frac{\Lambda'}{8\pi G'_N}$ are
almost compensated by the sum $\rho_{vac}^{(SM)} +
\rho_{vac}^{(SM')}$. This compensation is possible if we have the
dominance of DOF of fermions in the sum of zero-point energies of
all quantum field fluctuations. In the framework of the present
graviweak unification model with broken symmetry, the estimate
gives a cut-off scale, which is less than the Planck scale. If
this scale is equal to the supersymmetry breaking scale, then it
is extremely large: $M_{SUSY}\sim 10^{10}$ GeV, and not within the
reach of the LHC experiments. A possibility to reduce this value
of $M_{SUSY}$ is also discussed.

\section{Acknowledgments}

We thank Masud Chaichian and Tiberiu Harko for fruitful
discussions. L.V.~Laperashvili deeply thanks  A.~Garrett Lisi, also
O.V.~Kancheli, R.B.~Nevzorov,
V.A.~Novikov, M.A. Trusov and all members of the ITEP Theoretical
Seminar (Moscow, May, 16, 2013) for the interesting discussion and
advices. The support of the Academy of Finland under the Projects
No. 136539 and 140886 is gratefully acknowledged. CRD acknowledges
a scholarship from the Funda\c{c}\~{a}o para a Ci\^{e}ncia e a
Tecnologia (FCT, Portugal) (ref. SFRH/BPD/41091/2007). This work
was partially supported by FCT through the projects
CERN/FP/123580/2011 PTDC/FIS/ 098188/2008 and CFTP-FCT Unit 777
which are partially funded through POCTI (FEDER).


\begin{thebibliography}{99}
\bibitem{1a}
D.L.~Bennett, L.V.~Laperashvili, H.B.~Nielsen and A.~Tureanu, Int. J. Mod.
Phys. A {\bf 28}, 1350035 (2013),  arXiv:1206.3497.
\bibitem{1}
A.~Garrett~Lisi, L.~Smolin and S.~Speziale, J. Phys. A {\bf 43},
445401 (2010), arXiv:1004.4866.
\bibitem{2}
T.D.~Lee and C.N.~Yang, Phys. Rev. {\bf 104}, 254 (1956).
\bibitem{3}
I.Yu.~Kobzarev, L.B.~Okun and I.Ya.~Pomeranchuk, Yad. Fiz. {\bf
3}, 1154 (1966) [Sov. J. Nucl. Phys. {\bf 3}, 837 (1966)].
\bibitem{4}
K.~Nishijima and M.H.~Saffouri, Phys. Rev. Lett. {\bf 14}, 205
(1965).
\bibitem{5}  E.W.~Kolb, D.~Seckel and M.S.~Turner,
Nature {\bf 314}, 415 (1985).
\bibitem{6}
E.W.~Kolb, D.~Seckel and M.S.~Turner, Report
Fermilab-Pub-85/16-A (1985).
\bibitem{7}
Z.~Berezhiani, A.~Dolgov and R.N.~Mohapatra, Phys.Lett. B {\bf
375}, 26 (1996), hep-ph/9511221.
\bibitem{8}
Z.~Berezhiani, {\it Through the looking-glass: Alice's adventures
in mirror world}, in: Ian Kogan Memorial Collection ``From Fields
to Strings: Circumnavigating Theoretical Physics'', Eds.
M.~Shifman et al., World Scientific, Singapore, Vol.~3, pp.
2147-2195, 2005, hep-ph/0508233.
\bibitem{9}
R.~Foot, H.~Lew, and R.R.~Volkas,  Phys. Lett. B {\bf 272}, 67
(1991).
\bibitem{10}
R.~Foot, Mod. Phys. Lett. A {\bf 9}, 169 (1994), hep-ph/9402241.
\bibitem{11}
R.~Foot, Int. J. Mod. Phys. D {\bf 13}, 2161 (2004), astro-ph/0407623.
\bibitem{11a}
S.I.~Blinnikov and M.Yu.~Khlopov, Sov. Astron. {\bf 27}, 371
(1983) [Astron. Zh. {\bf 60}, 632 (1983)].
\bibitem{12}
L.B.~Okun, Phys. Usp. {\bf 50}, 380 (2007), hep-ph/0606202.
\bibitem{13}
S.I.~Blinnikov, Phys. Atom. Nucl. {\bf 73}, 593 (2010), arXiv:0904.3609.
\bibitem{14}
P.~Ciarcelluti, Int. J. Mod. Phys. D {\bf 19}, 2151 (2010),
arXiv:1102.5530, and references therein.
\bibitem{15}
Jian-Wei~Cui, Hong-Jian~He,  Lan-Chun Lu and Fu-Rong Yin, Phys. Rev. D {\bf 85}, 096003 (2012),
arXiv:1110.6893 [hep-ph].
\bibitem{17}
J.F.~Plebanski, J. Math. Phys. {\bf 18}, 2511 (1977).
\bibitem{18}
A.~Ashtekar, Phys. Rev. Lett. {\bf 57}, 2244 (1986).
\bibitem{19}
A.~Ashtekar, Phys. Rev. D {\bf 36}, 1587 (1987).
\bibitem{20}
T.~Jacobson and L.~Smolin, Phys. Let. B {\bf 196}, 39 (1987).
\bibitem{21}
R.~Capovilla, T.~Jacobson, J.~Dell and L.J.~Mason, Class. Quant. Grav. {\bf 8}, 41 (1991).
\bibitem{22}
R.~Capovilla, T.~Jacobson and J.~Dell, Class. Quant. Grav. {\bf 8}, 59 (1991).
\bibitem{23}
S.~Alexander, {\it Isogravity: Toward an Electroweak and Gravitational Unification},
arXiv:0706.4481.
\bibitem{24} 
F.~Nesti, Eur. Phys. J. C {\bf 59}, 723 (2009), arXiv:0706.3304.
\bibitem{25} 
F.~Nesti and R.~Percacci,  J. Phys. A
{\bf 41}, 075405 (2008), arXiv:0706.3307.
\bibitem{26} 
A.G.~Lisi, {\it An Exceptionally Simple Theory of Everything},
arXiv:0711.0770.
\bibitem{27}  
F.~Nesti and R.~Percacci, Phys. Rev. D {\bf 81}, 025010 (2010),
arXiv:0909.4537.
\bibitem{28} 
A.~Torres-Gomez and K.~Krasnov, Phys. Rev. D {\bf 81}, 085003 (2010), arXiv:0911.3793.
\bibitem{29} 
L.~Smolin, Phys. Rev. D {\bf 80}, 124017
(2009), arXiv:0712.0977.
\bibitem{30} 
K.~Krasnov, Gen. Rel. Grav. {\bf 43}, 1 (2011),
arXiv:0904.0423.
\bibitem{31} 
K.~Krasnov, Class. Quant. Grav. {\bf 26}, 055002 (2009),
arXiv:0811.3147.
\bibitem{32} 
K.~Krasnov, Class. Quant. Grav. {\bf 25}, 025001 (2008), gr-qc/0703002.
\bibitem{33} 
K.~Krasnov, Mod. Phys. Lett. A {\bf 22}, 3013 (2007),
arXiv:0711.0697.
\bibitem{34} 
M.P.~Reisenberger, Class. Quant. Grav.
{\bf 16}, 1357 (1999), gr-qc/9804061.
\bibitem{35} 
Eyo Eyo Ita III, {\it CDJ formulation from the instanton representation of
Plebanski gravity}, arXiv:0911.0604.
\bibitem{36} 
E.~Buffenoir, M.~Henneaux, K.~Noui and Ph.~Roche, Class. Quant.
Grav. {\bf 21}, 5203 (2004), gr-qc/0404041.
\bibitem{37} 
F.~Tennie and M.N.R.~Wohlfarth, Phys. Rev. D {\bf 82}, 104052
(2010), arXiv:1009.5595 [gr-qc].
\bibitem{38} 
D.L.~Bennett, C.R.~Das, L.V.~Laperashvili and H.B.~Nielsen, Int. J. Mod.
Phys. A {\bf 28}, 1350044 (2013), arXiv:1209.2155.
\bibitem{39} 
S.~Alexander, A.~Marciano and L.~Smolin, {\it Gravitational origin of
the weak interaction's chirality}, arXiv:1212.5246.
\bibitem{40} 
C.R.~Das and L.V.~Laperashvili, Phys. Atom. Nucl. {\bf 72}, 377
(2009), [Yad.Fiz. {\bf 72}, 407 (2009)].
\bibitem{41} 
C.R.~Das and L.V.~Laperashvili, Int. J. Mod. Phys. A {\bf 23}, 1863
(2008), arXiv:0712.1326.
\bibitem{42} 
C.R.~Das, L.V.~Laperashvili and A.~Tureanu, Eur. Phys. J. C {\bf
66}, 307 (2010), arXiv:0902.4874.
\bibitem{43}   
C.R.~Das, L.V.~Laperashvili and A.~Tureanu, AIP Conf. Proc. {\bf
1241}, 639 (2010), arXiv:0910.1669.
\bibitem{44}   
C.R.~Das, L.V.~Laperashvili and A.~Tureanu, Phys. Part. Nucl. {\bf
41}, 965 (2010), arXiv:1012.0624.
\bibitem{45}    
C.R.~Das, L.V.~Laperashvili, H.B.~Nielsen and A.~Tureanu, Phys.
Rev. D {\bf 84}, 063510 (2011), arXiv:1101.4558.
\bibitem{46}    
K.~Nakamura {\it et al.} (Particle Data Group), J. Phys. G {\bf
37}, 075021 (2010).
\bibitem{47}    
A.~Riess et al., Astrophys. J. Suppl. {\bf 183}, 109 (2009),
arXiv:0905.0697.
\bibitem{48}      
W.L.~Freedman et al., Astrophys. J. {\bf 704}, 1036 (2009),
arXiv:0907.4524.
\bibitem{49}
Z.~Berezhiani, Int. J. Mod. Phys. A {\bf 19}, 3775 (2004),
hep-ph/0312335.
\bibitem{50a}
L.~Bento and Z.~Berezhiani, Phys. Rev. Lett. {\bf 87}, 231304
(2001), hep-ph/0107281;
\bibitem{50b}
L.~Bento and Z.~Berezhiani, {\it Baryogenesis: The Lepton leaking
mechanism}, hep-ph/0111116.
\bibitem{50c}
L.~Bento and Z.~Berezhiani, Fortsch. Phys. {\bf 50}, 489 (2002).
\bibitem{51}
C.R.~Das, L.V.~Laperashvili, H.B.~Nielsen and A.~Tureanu, Phys.
Lett. B {\bf 696}, 138 (2011), arXiv:1010.2744.
\bibitem{52}
M.~Fukugita and T.~Yanagida, Phys. Lett. B {\bf 174}, 45 (1986).
\bibitem{53}  
W.~Buchmuller, R.D.~Peccei and T.~Yanagida, Ann. Rev. Nucl. Part.
Sci. {\bf 55}, 311 (2005), arXiv:hep-ph/0502169.
\bibitem{54}
E.K.~Akhmedov, Z.G.~Berezhiani and G.~Senjanovic, Phys. Rev. Lett.
{\bf 69}, 3013 (1992), hep-ph/9205230.
\bibitem{55}
Z.~Berezhiani, D.~Comelli and N.~Tetradis, Phys. Lett. B {\bf
431}, 286 (1998), hep-ph/9803498.
\bibitem{55a}
Z.~Berezhiani, P.~Ciarcelluti, D.~Comelli and F.L.~Villante, Int. J.
Mod. Phys. D {\bf 14}, 107 (2005), astro-ph/0312605.
\bibitem{55b}
Z.~Berezhiani, L.~Kaufmann, P.~Panci, N.~Rossi, A.~Rubbia and
A.~Sakharov, {\it Strongly interacting mirror dark matter},
CERN-PH-TH-2008-108 (May 2008).
\bibitem{56}
Z.~Berezhiani and R. N.~Mohapatra, Phys. Rev. D {\bf 52}, 6607
(1995), hep-ph/9505385.
\bibitem{56a}
R.~Foot, H.~Lew and R.R.~Volkas, Mod. Phys. Lett. A {\bf 7}, 2567
(1992).
\bibitem{56b}
R.~Foot and R.R.~Volkas, Phys. Rev. D {\bf 52}, 6595 (1995),
hep-ph/9505359.
\bibitem{57}
E.W.~Mielke, Phys. Rev. D {\bf 77}, 084020 (2008), arXiv:0707.3466.
\bibitem{58}
G.~de~Berredo-Peixoto and I.L.~Shapiro, Phys. Rev. D {\bf 70}, 044024
(2004), hep-th/0307030.
\bibitem{59}
L.V.~Laperashvili, Phys. Atom. Nucl. {\bf 57}, 471 (1994)
[Yad.Fiz. {\bf 57}, 501 (1994)]
\bibitem{60}  
D.L.~Bennett, L.V.~Laperashvili and H.B.~Nielsen, {\it What Comes
Beyond the Standard Models?} Proceedings, 9th Workshop, Bled,
Slovenia, September 16-27, 2006 (DMFA, Zaloznistvo, Ljubljana,
2006), hep-ph/0612250.
\bibitem{60a}   
D.L.~Bennett, L.V.~Laperashvili and H.B.~Nielsen, {\it What Comes
Beyond the Standard Models?}, Proceedings, 10th Workshop, Bled,
Slovenia, July 17-27, 2007 (DMFA, Zaloznistvo, Ljubljana, 2007),
arXiv:0711.4681.
\bibitem{61}
N.~Yunes, R.O'~Shaughnessy, B.J.~Owen and S.~Alexander, Phys. Rev.
D {\bf 82}, 064017 (2010), arXiv:1005.3310.
\bibitem{62}
R.~Foot, A.~Kobakhidze and R.R.~Volkas, Phys. Rev. D {\bf 84}, 09503
(2011), arXiv:1109.0919.
\bibitem{63}
Z.~Berezhiani, L.~Pilo and N.~Rossi, Eur. Phys. J. C {\bf 70},
305 (2010), arXiv:0902.0146.
\bibitem{64}
R.N.~Mohapatra and V.L.~Teplitz, Phys.~Lett. B {\bf 462}, 302
(1999), astro-ph/9902085.
\bibitem{65}
S.L.~Glashow, Phys. Lett. B {\bf 167}, 35 (1986).
\bibitem{66}
A.~Badertscher et al., Int. J. Mod. Phys. A {\bf 19}, 3833 (2004), hep-ex/0311031.
\bibitem{66a}
Z.~Berezhiani and A.~Lepidi, Phys. Lett. B {\bf 681}, 276 (2009),
arXiv:0810.1317.
\bibitem{67}
S.V.~Demidov, D.S.~Gorbunov and A.A.~Tokareva, Phys. Rev. D {\bf
85}, 015022 (2012), arXiv:1111.1072.
\bibitem{68}
Z.~Berezhiani and L.~Bento, Phys. Rev. Lett. {\bf 96}, 081801
(2006), hep-ph/0507031.
\bibitem{69}
Z.~Berezhiani and L.~Bento, Phys. Lett. B {\bf 635}, 253 (2006), hep-ph/0602227.
\bibitem{70}
R.N.~Mohapatra, S.~Nasri and S.~Nussinov, Phys. Lett. B {\bf 627},
124 (2005), hep-ph/0508109.
\bibitem{71}
Yu.N.~Pokotilovski, Phys. Lett. B {\bf 639}, 214 (2006),
nucl-ex/0601017.
\bibitem{72}
Z.K.~Silagadze, Phys. Atom. Nucl. {\bf 60}, 272 (1997) [Yad. Fiz.
{\bf 60}, 336 (1997)], hep-ph/9503481.
\bibitem{73}
V.~Berezinsky and A.~Vilenkin, Phys. Rev. D {\bf 62}, 083512
(2000), hep-ph/9908257.
\bibitem{74}
A.Yu.~Ignatiev and R.R.~Volkas, Phys. Lett. B {\bf 487}, 294 (2000), hep-ph/0005238.
\end{thebibliography}
\end{document}